\documentclass[12pt,preprint]{aastex}
\usepackage{natbib}
\begin{document}
\title{The Hot Components of AM CVn Helium Cataclysmics}
\author{Edward M. Sion}
\affil{Dept. of Astronomy and Astrophysics,
Villanova University,
Villanova, PA 19085,
email: edward.sion@villanova.edu}

\author{Albert P.Linnell}
\affil{Dept. of Astronomy,
University of Washington,
Seattle, WA 98195,
email: linnell@astro.washington.edu}

\author{Patrick Godon}
\affil{Dept. of Astronomy and Astrophysics,
Villanova University,
Villanova, PA 19085,
email: patrick.godon@villanova.edu}

\author{Ronald-Louis Ballouz}
\affil{Dept. of Astronomy and Astrophysics,
Villanova University,
Villanova, PA 19085,
email: ronald-louis.ballouz@villanova.edu}

\begin{abstract}
We present the results of a multi-component synthetic spectral analysis of the archival far ultraviolet spectra of the hot components of several AM CVn double degenerate interacting binaries with known distances from trigonometric parallaxes. Our analysis was carried out using the code BINSYN (Linnell \& Hubeny 1996) which takes into account the donor companion star, the shock front which forms at the disk edge and the FUV and NUV energy distribution. We fixed the distance of each system at its parallax-derived value and adopted appropriate values of orbital inclination and white dwarf mass. We find that the accretion-heated "DO/DB" WDs are contributing significantly to the FUV flux in four of the systems (ES Ceti, CR Boo, V803 Cen, HP Lib, GP Com). In two of the systems, GP Com and ES Ceti, the WD dominates the FUV/NUV flux. We present model-derived accretion rates which agree with the low end of the range of accretion rates derived earlier from black body fits over the entire spectral energy distribution. We find that the WD in ES Ceti is very likely not a direct impact accretor but has a small disk. The WD in ES Ceti has $T_{eff} \sim 40,000 \pm 10,000$K. This is far cooler than the previous estimate of Espaillat et al.(2005). We find that the WD in GP Com has $T_{eff} = 14,800 \pm500$K, which is hotter than the previously estimated temperature of 11,000K. We present a comparison between our empirical results and current theoretical predictions for these systems.
\end{abstract}

Subject Headings: Stars: cataclysmic variables, white dwarfs, Physical
Processes: accretion, accretion disks

\section{Introduction}

A puzzling and exotic subset of all CVs are the nearly pure helium systems known 
as the AM CVn binaries.They have the shortest orbital periods 
(5min-1h) known for any interacting binary and hence are the most compact of any known interacting systems.Their spectra are dominated by helium spectral features arising from a helium accretion disk which has formed via mass transfer from a degenerate or
semi-degenerate helium donor star filling its Roche lobe (Paczynski 1967; Deloye \& Bildsten 2007). 
The basic model was originally proposed by Paczynski (1967), 
and Faulkner, Flannery, and Warner (1972) with their binary nature 
first confirmed by Nather, Robinson, and Stover (1981). 
The mass transfer is driven by angular momentum loss due to 
gravitational wave radiation (GWR). Because GWR depends very strongly on the orbital period,
this suggests a rapidly dropping mass-transfer rate as a function of orbital period.
However, this hypothesis has never been tested with either actual realistic helium 
accretion disk models with vertical structure, or with hot helium-rich, high gravity photospheres or combinations of photospheres and disks. Indeed much of the effort
has focussed only upon the prototype system AM CVn itself (Nasser and Solheim 2001) which is always observed in its high state when the helium accretion disk is very luminous.

The spectroscopic, orbital and physical properties of these systems are comprehensively reviewed by Warner (1995), and Solheim (2010). 
There are three types of AM CVn systems. Objects like AM CVn and HP Lib
are in continuous high states and show apparent He disk spectra with absorption lines.
Four objects have both outburst states and quiescence states as in dwarf novae (e.g. CR Boo, V803 Cen), and three objects are seen in continuous low states (e.g. GP Com).
In their bright states, the AM CVns 
spectroscopically resemble the prototype AM CVn in its continual bright
state in which the absorption lines of an optically thick helium disk and
wind dominate their optical and FUV spectra (e.g., Groot et al. 2001 and
references therein). It is unclear whether the helium accretion disk also dominates the 
light during the low states.

The AM CVn objects are fundamentally important because of their exotic evolutionary history which may have involved double common envelope evolution and led to their 
nearly pure helium composition in which the physics of helium accretion under extreme conditions can be explored.
Most importantly, AM CVn systems may be a significant channel for the production of Type Ia
supernovae and can contribute up to 25\% of the galactic Type Ia
supernova rate (Nelemans et al 2001). While some population synthesis caclulations
indicate they may not be a significant SNIa progenitor channel (Yungelson and Solheim 2002), a recent
discovery of the progenitor of the type Ia supernova SN2007 as an AM CVn star
strongly suggests that indeed some AM CVn stars may be Type Ia progenitors (Varus \& Nelemans
2008). Further support from recent theoretical work suggests that AM CVn systems
could indeed produce Type Ia SN ( Bildsten et al.2007). Interestingly, the final evolutionary state of these systems, if they don't explode as SN Ia when mass transfer has ceased, 
may be a DB WD (Nather 1985) with a planet or brown dwarf as companion (Gonzalez Perez
2004). Furthermore, as their mass transfer is driven by
General Relativity, and the AM CVns are predicted to be a significant
source of the low frequency gravitational radiation background (Warner
1995), they are currently the only known sources for the space-based GWR detector
Laser Interferometer Space Antenna (LISA) since this NASA/ESA LISA mission will be sensitive to only these relatively rare short period systems.

One major obstacle to synthetic spectral analysis has been the large uncertainty in their distances. Since trigonometric parallaxes have now been measured with the Hubble FGS for 
several AM CVn systems (Roelofs et al. 2007), this makes it possible to constrain the parameter space of model fits and attempt to disentangle the flux contributions of 
multiple emitting components. In this work, we exploit this newly available distance information to carry out extensive multi-component model fitting to the FUV/NUV spectra of AM CVn systems in high states and in low states, in much the same way we have done for H-rich CVs (Sion et al. 2008, Urban and Sion 2006). In this paper we confine our attention to five systems for which the distances are known from the parallax measurements of Roelofs et al. (2007) and for which archival IUE spectra exist.

The reddening of the systems was determined based upon all estimates 
listed in the literature. The three principal sources of reddening were
the compilations of Verbunt (1987), laDous (1991) and Bruch \& Engel
(1994) or from the Galex archive on MAST. The spectra were de-reddened with 
the IUERDAF IDL routine UNRED.

\clearpage 

In table 1, we tabulate the published values from Solheim (2010), Roelofs et al.(2007a) 
and Bildsten et al. (2006) and Nelemans (2005) and references therein, listed by column, (1) the system name,
(2) the orbital period $P_{orb}$ (s), 
(3) the white dwarf mass (solar masses), 
(4) the secondary mass  (solar masses), 
(5) the orbital inclination $i$ (degrees), 
(6) the reddening E(B-V), and (7) the distance in parsecs.

\begin{center}
{\bf
Table 1. Summary of Observational Properties of AM CVn stars
}
\begin{small}
\begin{tabular}{lcccccc}
\hline
Name     &$P_{orb}$& $M_{wd}$     &   $M_2$        &  i     & E(B-V) &  d     \\
         &  (s)    & ($M_{\odot})$& $(M_{\odot})$  & (deg)  &        & (pc)   \\
\hline
ES Cet   & 621     & $<$ 0.8    & -  & -      & 0.025 &  350 \\ 

GP Com   & 2794    & 0.5 - 0.68  & 0.009 - 0.012 & -      & 0.026 &  75  \\

HP Lib   & 1103    & 0.49 - 0.80 & 0.048 - 0.088 & 26 -34 & 0.149 & 197  \\

CR Boo   & 1471    & 0.67 - 1.10 & 0.044 - 0.088 & 30     & 0.023 & 337 \\

V803 Cen & 1612    &0.78 - 1.17  & 0.059 - 0.109 & 12- 15 & 0.120 & 347 \\
\hline
\end{tabular}
\end{small}
\end{center}

\section{Far Ultraviolet Archival Spectra}

All the spectral data were obtained from the Multimission Archive at Space Telescope (MAST). 
We restricted  our selection to those systems with IUE SWP spectra, with resolution of 5\AA\ and a spectral 
range of 1170\AA\ to 2000\AA.  After downloading all of the IUE spectra; we re-calibrated 
them according to the Massa and Fitzpatrick flux calibration correction, deleted bad pixels 
and saved the spectra for plotting. All spectra were taken through the large aperture at 
low dispersion. 

In Table 2, an observing log of the IUE archival spectra is presented by column: (1) the 
system name, (2) the SWP and paired LWP spectrum number, (3) exposure time in seconds, 
(4)dispersion, (5) the aperture diameter, (6) the date and time of the observation, and 
(7) the brightness state of the system when the spectrum was obtained. 

\begin{deluxetable}{lcccccc}  
\tablecaption{IUE Observing Log}
\tablenum{2}
\tablewidth{0pc}
\tablecolumns{8}
\tablehead{
\colhead{System} &\colhead{Data ID} &\colhead{t$_{exp}$}&\colhead{Disp.}&\colhead{Ap.}&\colhead{Date of Obs.} &\colhead{State}   \\ 
\colhead{Name} &                  &\colhead{(s)}        &   &  &\colhead{yyyy-mm-dd hh:mm:ss} &            
}
 \startdata        
ES Ceti & SWP30020 & 3600  & LOW & LARGE & 1987-01-04 21:46:25 & high \\
GP Com  & SWP46900 & 19500 & LOW & LARGE & 1993-02-08 06:04:06 & low \\ 
        & LWP20586 & 14400 & LOW & LARGE & 1991-06-13 00:20:25 & low \\ 
HP Lib  & SWP51283 & 3000  & LOW & LARGE & 1994-07-01 20:38:37 & high \\ 
        & LWP28520 & 2700  & LOW & LARGE & 1994-07-01 23:12:25 & high \\ 
CR Boo  & SWP33087 & 25200 & LOW & LARGE & 1988-03-13 12:56:17 & low  \\
        & SWP33086 & 37200 & LOW & LARGE & 1988-03-13 01:45:44 & low \\ 
        & SWP33077 & 3600  & LOW & LARGE & 1988-03-10 20:10:40 & high  \\ 
         & LWP12836 & 3600  & LOW & LARGE & 1988-03-10 21:16:53 & high  \\ 
        & LWP12847 & 5400  & LOW & LARGE & 1988-03-13 00:08:50 & low  \\ 
V803 Cen& SWP44085 & 7200  & LOW & LARGE & 1992-03-01 05:03:16 & low \\ 
        & LWP22482 & 3600  & LOW & LARGE & 1992-03-01 03:58:46 & low \\ 
        & SWP38270 & 5880  & LOW & LARGE & 1990-02-28 10:32:21 & high  \\
        & LWP17437 & 2400  & LOW & LARGE & 1990-02-28 09:45:58 & high \\ 
\hline
\enddata
\end{deluxetable}          

Since AM CVn systems exhibit considerable variation in brightness (with some systems 
undergoing outbursts), it is essential that the selection of IUE spectra for analysis 
sample the maximum and minimum brightness states of a given system. 

The systems ES Ceti and HP Lib are always seen in the high brightness 
state (i.e. "permanent outburst" by analogy with the UX UMa subtype of nova-likes among the H-rich cataclysmics). Thus, in Table 2 their spectra correspond to a "high state" and are designated as such column 7 of Table 2.

GP Com on the other hand is always in a low brightness state and has never been observed to be in a higher brightness state (i.e. "outburst"). Hence its spectra were obtained in the low state and designated as such in Column 7 of Table 2.   

Both CR Boo and V803 Cen have both high states and low states, by analogy with dwarf novae and VY Scl-type nova-like variables among H-rich CVs. Hence, we have used the AAVSO and VSNET light curve data on both of these systems as well as the confirmatory flux levels of each SWP + LWP matching pair of spectra to ascertain whether they correspond to a high state or low state.

For CR Boo, the IUE spectra SWP33087 and LWP 12847 were obtained during 
a low state (quiescence) of CR Boo on 17 May 1984 when the flux level 
(at 1280\AA) was  $\sim 1.5 \times 10^{-14}$ergs/cm$^2$/s/\AA . 
Hence, the designation in Column 7 of Table 2 refers to the low state. 
This is also mentioned in Wood et al.(1987). The IUE spectral pair 
SWP33077 + LWP12836 were obtained on 3 March 1988 when the flux level 
was higher ( $\sim 5.0 \times 10^{-14}$ergs/cm$^2$/s/\AA , at 1280\AA ) 
and CR Boo was in its high brightness state. Hence, the designation in 
Column 7 of Table 2 refers to the high ("outburst") state. 

For V803 Cen, the IUE spectra SWP44085 + LWP 22482 were obtained during
a low brightness state of the system when the flux level 
at 1280\AA\ was $\sim 7.5 \times 10^{-14}$ergs/cm$^2$/s/\AA , 
hence they are designated as low state spectra 
in column 7 of Table 2. The IUE spectral pair SWP38270 + 
LWP17437 were obtained during a high brightness state of V803 Cen 
when the flux level 
at 1280\AA\ was $\sim 2.5 \times 10^{-13}$ergs/cm$^2$/s/\AA , 
hence they are designated as high state spectra in column 7 of Table 2.

\section{Synthetic Spectral Analysis}

We combined He-rich accretion disk models and He-rich white dwarf 
models constructed with BINSYN, a comprehensive software package 
for simulating binary stars with or without accretion disks.
 In 
the case of accretion disk systems (Linnell \& Hubeny 1996) the 
program constructs a generalized accretion disk with an assignable 
radial temperature gradient; the standard model case with $T_{eff}(r)$ 
proportional to $r^{-3/4}$ is a special case. The program calculates 
a synthetic spectrum of the primary star, the secondary star,
the accretion disk face, the accretion disk rim, and the entire 
system at arbitrary orbital longitudes and orbital inclinations. 
The program allows for eclipse effects, mutual irradiation, and a 
hot spot (shock front) at the disk edge. The package also calculates 
light curves by synthetic photometry.

In all synthetic spectra reported here the adopted number ratio He/H was $1.0 \times 10^4$.
Since the spectra of stellar models closely approximate accretion disk annuli
spectra of the same $T_{eff}$ (Hubeny(1990)) we have used the same set
of synthetic spectra for both the WD and accretion disk annuli.
The radii of the white dwarf models for a given input value of $\log{g}$ 
and $\log{T_{eff}}$ were read from Fig.4a in Panei et al. (2004).
Using our synthetic spectra and successive trial parameters we obtained
the five AM CVn system parameters listed in Table 1. The best-fitting combination 
of models was selected on the basis of consistency with the trigonometric parallax
distances of Roelofs et al. (2007a), and visual estimates of the goodness of fit to the continuum slope and to any absorption features.
Some parameters such as the white dwarf mass are not precisely known.
Since the white dwarf radius is fixed by its mass and the radius
enters into the scale factor of a given fit, the white dwarf contribution to the synthetic spectrum is affected appreciably.
For our study, we have adopted values of the white dwarf mass for each system which are published in the 
literature. We discuss each system in the subsections below.

\subsection{ES Ceti}

ES Ceti seems to remain in the same brightness state (a high state)
by analogy with nova-like variables and has only one 
SWP spectrum (SWP30020) and no LWP spectrum. Thus, our wavelength coverage is not as wide as the other sources in this study.
At its very short $P_{orb}$, ES Ceti has been considered a likely direct impact accretor. As we show below, in our model of the system, ES Ceti is not a direct impact accretor and an accretion disk is necessary to understand the FUV observational data.
We have adopted a WD mass of $0.7 M_{\odot}$ (Solheim 2010 and references therein). Assuming the
low amplitude light variation (Espaillat et al. 2005) represents only the phase variation
from the stream impact, not an eclipse, we have adopted a low inclination of 41 degrees.
Separate tests with a range of helium-rich WD $T_{eff}$ spectra, scaled to fit the SWP spectrum, established a source $T_{eff}$ close to 40,000K. A nominal synchronously-rotating $0.7 M_{\odot}$ 40,000K WD, without an accretion disk, must be at a distance
of 213pc to fit the observed spectrum, in disagreement with our adopted distance of 350 pc from Solheim (2010). In order to achieve better agreement with our adopted distance, we tried a standard accretion disk model. In Fig.1, we display a standard model helium accretion disk with 33 annuli (rings), an outer radius of $0.03R_{\odot}$ and an accretion rate $5 \times 10^{-10} M_{\odot}$/yr (upper synthetic spectrum) and the WD contribution to the system spectrum (lower synthetic spectrum). 

The calculated flux from both the WD and the system synthetic spectrum has been scaled by 1.16639E29, corresponding (exactly) to the distance 
of 350pc (Solheim 2010). Note that the accretion disk flux lies significantly below the observed continuum level. In judging the quality of fit, we have used eye estimates of 1/5 of the peak-to-peak variation in the observed continuum spectrum, excepting emission lines. 
The synthetic spectrum displacement below the observed spectrum can be ameliorated by either decreasing the WD mass (making the WD larger)
or increasing the accretion rate. We tried a hotter disk by increasing the accretion rate to $2.5 \times 10^{-9} M_{\odot}$/yr and $5 \times 10^{-8} M_{\odot}$/yr. In the former case, 
the disk spectrum still looked too cool while in the latter case, the disk flux was too high relative to the observations. Increasing the outer disk radius to $R_{disk} = 0.035$ resulted in excess flux, based on our eye estimate criterion. Our optimal fit resulted when we 
increased the disk radius only slightly to $R_{disk} = 0.032 R_{\odot}$. This final fitting result is displayed in Fig.2. Thus, we find that the FUV spectrum of ES Ceti is best represented by a combination of an (adopted) $0.7 M_{\odot}$
WD with $T_{eff} = 40,000$K and an accretion disk dominating the FUV light with $\dot{M} = 2.5 \times 10^{-9} M_{\odot}$/yr. The empirically-determined outer accretion disk radius
is much smaller than the tidal truncation radius (Warner 1995).
Fig.3 shows the orbital plane view of the system with the accretion disk marked by the diagonal line region. The mass transfer stream is shown, terminating on the accretion disk rim. Continuation of
the stream shows that, in the absence of an accretion disk, the stream would miss the WD if one neglects stream spreading (Lubow and Shu 1975, 1976). 

\subsection{GP Com} 

GP Com is always seen in the same brightness state (a low state by analogy with low states of H-rich nova-like variables and dwarf novae during quiescence). 
The pair of IUE spectra that we selected for analysis are SWP46900 
and LWP20586. As with ES Ceti, initial tests established a source $T_{eff}$=14,800K.
Standard model accretion disks with median $T_{eff}$ values 
around 14,800K have $\dot{M}$ values around $3-4 \times 10^{-11} M_{\odot}$/yr.
We were constrained in this system analysis by the lack of convergence of
tlusty for $T_{eff}$ values below 14,000K. Nather suggests a low mass for the WD
(Nather et al. (1981)). The published range of white dwarf masses is between 0.5 and $0.7 M_{\odot}$ (see Table 1). Following Marsh (1999), we adopted $M_{\odot}$=0.7. Marsh (1999) shows that the
inclination is large; we adopt $i=75$ deg.
A standard model accretion disk with the
$\dot{M}$ just described requires annulus $T_{eff}$ values down to 11,000K.
Consequently we have forced the accretion disk to be isothermal at 14,800K.
In Fig.4, we display the best fitting WD + disk model, with the 14,800K WD 
seen at the bottom of the figure. The best synthetic spectral fit to 
the outer radius of the annulus is $0.030R_{\odot}$. 
The accretion disk is isothermal with $T_{eff}=14,800$K and the WD has the same 
$T_{eff}$. The accretion disk radius is $0.03R_{\odot}$ while the WD radius is 
$0.011R_{\odot}$. The WD contribution is about 1/2 the system flux or a bit 
less. The uncertainty in the WD temperature is $\pm 1500$K.  

\subsection{HP Lib}    

HP Lib is also continually seen in the same brightness state, a high state with no known low state ever having been recorded. Thus, it was reasonable to try standard, 
steady state, helium accretion disk fits with standard model radial 
temperature gradients. As with previous systems, we first determined
the source $T_{eff}$ and found it to be 30,000K. There are no eclipses
and we adopted $i=30$ deg, a value in the middle of the range of published inclinations given in Table 1.
Trying a standard model accretion disk, we obtained a fairly successful fit to the 
IUE spectra SWP51283 + LWP28520. 
The $\dot{M}$ is $8.0 \times 10^{-10}M_{\odot}$/yr. 
The temperatures of the disk annuli range from 31,076K to 22,304K. 
In order to fit the observed fluxes, it was necessary to adjust the outer radius of the accretion disk. We tried outer radii of $0.0370R_{\odot}$, 
$0.0400R_{\odot}$, and $0.038R_{\odot}$. The best-fitting result was 
with $R_{disk} = 0.038 R_{\odot}$ and is shown in Fig.5.
Note the small WD contribution (bottom plot) to the system synthetic spectrum.

\subsection{CR Boo}

CR Boo, like V803 Cen and CP Eri, has both high and low optical 
brightness states in analogy with H-rich dwarf novae and nova-like variables. We adopted $M_{wd} = 0.9M_{\odot}$ (middle of the mass range in Roelofs et al.2007a), inclination 
$i=30^{\circ}$ (Nasser et al. 2001), the WD radius from
Table 4a in Panei et al. (2000), $R_{wd} = \sim 0.095R_{\odot}$, 
and the mass of the secondary, $M_2 \sim 0.023M_{\odot}$ 
from Table 1 of Bildsten et al. (2006). 

The IUE spectra SWP33087 + LWP12847 were obtained during the low state of CR Boo. The LWP spectrum is noisy and was scaled by $-0.05 \times 10^{-14}$ to combine it
with the SWP spectrum. 
We first tried fitting WD synthetic spectra with $\log{g}=8.0$ 
and He/H=$1 \times 10^4$, and a range of $T_{eff}$ values to 
the combination swp33087+lwp12847. This was 
followed by standard accretion disk model fitting. 
The resulting best fits were obtained with a 
combination of an accretion disk and white dwarf photosphere. 
The outer radius of the accretion disk during the 
low state was $R_{disk} = 0.017 R_{\odot}$. 
Fig.6 shows the best fitting combination of an 
He-rich WD plus accretion disk model to the low state of CR Boo.  

The IUE spectra SWP33077+ LWP12836 were obtained during the high state of CR Boo. 
The difference in flux levels between the high and low states of CR Boo is not very large. 
The high state 
spectrum has a flux level of $\sim 5 \times 10^{-14}$ergs/cm$^2$/s/\AA\  at 1280\AA\  
compared with the low state flux level of $\sim 1.5 \times 10^{-14}$ergs/cm$^2$/s/\AA .  
In Fig.7, we display the best fitting combination of WD plus disk SWP33077+ LWP12836.
The only difference between Fig.6 and Fig.7 is the outer radius of the accretion disk. 
The outer radius of the accretion disk fit to the high state spectrum was 
$R_{disk} = 0.035 R_{\odot}$ while it was $R_{disk} = 0.017$ 
in the low state. The WD parameters and the derived accretion 
rate are exactly the same in both figures. The change in flux 
between the two figures is exclusively a result of the change 
in radius of the accretion disk. Thus, the fractional contribution 
of the WD is larger for the smaller accretion disk radius in Fig.5. 
Note that the solid black line toward the bottom of the plot is 
the contribution of the WD spectrum alone while the upper solid 
black line fitting the observed spectrum itself is the combined 
synthetic spectrum of the WD plus the accretion disk.
We determined $\dot{M}$ by running several BINSYN models, varying the 
adopted $\dot{M}$ until the annulus temperature extremes were about equally spaced around 30,000K. The temperature extremes are 34,795K and 23,000K with the former temperature characterizing only the very 
narrow annuli at the inner edge of the accretion disk.

\subsection{V803 Cen}    

For V803 Cen with its trigonmetric parallax distance, we adopted 
$M_{wd} =0.9 M_{\odot}$ (Roelofs et al. 2007a, b) and $i=15$ deg
for the model fitting. The IUE spectra SWP44085 + LWP22482 were obtained during the low state 
of V803 Cen. The flux level at 1280\AA is  
$\sim 7.5 \times 10^{-14}$ ergs/cm$^2$/s/\AA.
The best fit to the low state of V803 Cen was a combination of an 
accretion disk with an accretion disk outer radius of 
$0.030R_{\odot}$ and $\dot{M} = 5 \times 10^{-10}M_{\odot}$/yr 
and a WD with $T_{eff} = 30,000$K contributing only $\approx$10\% 
of the FUV light. The standard model accretion disk has a $T_{eff}$ 
range from 34,027K to 24,909K.
The fit with this model is fairly good but the best fit is with an isothermal 30,000K accretion disk with the same outer radius. 
A plot of this result for the low state of V803 Cen is displayed in Fig.8.

The IUE spectra SWP38270 + LWP17437 were obtained during the high state of 
V803 Cen. The flux level at 1280\AA is $2.5 \times 10^{-13}$ergs/cm$^2$/s/\AA. 
For this combination of spectra,
we found that by using an isothermal annulus with $T_{eff} = 30,000$K with an outer radius of 
$0.055 R_{\odot}$, we obtained a very good fit. 
On the other hand, when we used a full standard model accretion disk 
+ WD with an outer radius of $R_{disk} = 0.050$ and an accretion rate 
$\dot{M} = 5 \times 10^{-10} M_{\odot}$/yr, the fit was not as good 
as the isothermal ($T = 30,000$K) annulus case.  
A plot of the best-fit to the high state spectrum SWP38270 + LWP22482 of V803 Cen with an isothermal 30,000K annulus with an outer 
radius of $0.035R_{\odot}$, is displayed in Fig.9.

In Table 3, we summarize the best-fitting parameters of this selected 
sample of AM CVn systems where the entries by column are (1) the 
system name, (2) white dwarf mass, (3) inclination 
angle, (4) best-fitting model distance in pc, (5) $T_{WD}$, (6) $\dot{M}$, 
(7) Dominant FUV source. 

\clearpage

\begin{deluxetable}{lccccccc}
\tablecaption{AM CVn Fitting Results}
\tablenum{3}
\tablecolumns{8}
\tablewidth{0pc}
\tablehead{
\colhead{System}
&\colhead{M$_{wd}$}
&\colhead{\it{i}}
&\colhead{d$_{model}$}
&\colhead{$T_{wd}$} 
&\colhead{\.{M}}
&\colhead{Brightness State}
&\colhead{Dominant FUV Source}
\\                   
&\colhead{(M$_{\odot}$)}
&\colhead{($\deg$)}
&\colhead{(pc)}
&\colhead{(1000K)} 
&\colhead{(M$_{\odot}$yr$^{-1}$)}
&\colhead{     }
}
\startdata
ES Ceti & 0.7  & 41 & 350  & $40 \pm 10$  & $2.5\times10^{-9}$  & High & WD \\ 
GP Com  &  0.7 & 75 & 75   & $14 \pm 0.5$ & $3-4\times10^{-11}$ & Low & WD \\
HP Lib  & 0.7  & 30 & 197  &    -            & $8\times10^{-10}$ & High & He disk \\    
CR Boo  & 0.9  & 30 & 337  &    -         &$5\times10^{-10}$ & Low   & He disk \\
CR Boo  & 0.9  & 30 & 337  &    -        & $5\times10^{-10}$ & High   & He disk \\
V803 Cen & 0.9 & 15 & 347  &    -          & $5\times10^{-10}$ &  Low   & He disk\\
V803 Cen & 0.9 & 15 & 347  &    -          & $5\times10^{-10}$ &  High  & He disk \\
\hline
\enddata
\end{deluxetable} 

\clearpage

\section{Conclusions}

Our synthetic spectral analysis of the far ultraviolet spectra of AM CVn systems obtained with the IUE
spacecraft reveals a number of new findings. First, we have found that in four of the 
systems (ES Ceti, CR Boo, V803 Cen, HP Lib, GP Com), the accretion-heated "DO/DB" WDs are contributing significantly 
to the FUV flux. This confirms the prediction by Sion et al. (2006) and Bildsten et al. (2006)
that the accreting white dwarf is a significant contributor to the FUV but not in the optical (with the exception of GP Com).
Nevertheless, He-rich disks in the FUV with a range of temperatures dominate the FUV flux.  
In two of the systems, GP Com and ES Ceti, the WD dominates the FUV/NUV flux. 

Our estimated accretion rates derived in the FUV/NUV alone from disk and 
photosphere models, using the accurate trigonometric parallax distances, are in 
agreement with the lowest value of the range of accretion rates derived for HP Lib, CR Boo and V803 Cen 
from "black body" fits over the entire SED by Roelofs et al. (2007). The system with the highest derived accretion rate
is the shortest period system ES Ceti with 
$\dot{M}   = 2.5 \times 10^{-9} M_{\odot}$/yr 
while our lowest derived accretion rate
is for the longest period system, GP Com with 
$\dot{M}   = 3-4 \times 10^{-11} M_{\odot}$/yr.  

For HP Lib, which remains at the same brightness level, and the "outbursting" systems, CR Boo and V803 Cen,
all three of which have similar orbital periods, the derived accretion 
rates are also similar at $3 -5 \times 10^{-10} M_{\odot}$/yr. 
In the case of CR Boo and V803 Cen, the variation in the flux level between the high state and the low state is not large and 
can be accounted for by a relatively small change in the size of the accretion disk between the two states rather than by an
appreciable change in the accretion rate between the two states.

We have found that the WD in ES Ceti is very likely a direct impact accretor, 
but a small disk around the WD cannot be ruled out. The WD in ES 
Ceti has $T_{eff} \sim 40,000 \pm 10,000$K and dominates the systems's FUV flux. Our derived temperature is far cooler 
than the previous estimate of Espaillat et al.(2005). 
We find that the WD in GP Com has $T_{eff} = 14,800 \pm 500$K which is hotter
than the previously estimated temperature of 11,000K by Roelofs et al. 
(2007) while the accretion rate that we estimate, $3-4 \times 10^{-11}M_{\odot}$/yr 
is a factor of 10 higher than that derived by Roelofs et al.(2007).

Finally, given the relatively poor quality and limited sensitivity of the spectra
obtained with the small aperture IUE telescope, it is clear that further progress on
AM CVn systems requires higher quality spectra with HST, especially with the capabilities 
of COS to reach the Lyman Limit.

We thank an anonymous referee for helpful comments.
This research was supported by NASA ADP grant NN09AC94G
and in part by NSF grant AST0807892, both to Villanova University

\section{References} 
\noindent 
Bildsten, L., et al. 2008, ApJ, 662, 95 \\
Bildsten et al.2006, ApJ, 640, 465 \\
Bruch, A., \& Engel, A. 1994, A\&A, 104, 79 \\
Cropper M., Harrop-Allin M.K., Mason K.O., et al., 1998, MNRAS, 293, L57 \\
Deloye C.J., Bildsten L., 2003, ApJ, 598, 1217 \\
Deloye, C.J.,et al.2007, MNRAS, 381, 525\\
Espaillat, C., Patterson, J., Warner, B., Woudt, P. 2005, PASP, 117, 189\\
Faulkner, J., Flannery, B.P., Warner, B.,1972, ApJ, 175,L79 \\
Groot, P., et al. 2001, ApJ, 558, 123 \\
Israel G., et al., 2004, in Tovmassian G., Sion E., eds., Compact binaries in The Galaxy
   and beyond, volume 20 of RevMexAA (SC), p. 275 \\
Hubeny, I. 1990,ApJ, 351, 632\\
Linnell A.P., Hubeny, I., 1996, ApJ, 471, 958 \\
Lubow, S.H., Shu, F.H. 1975, ApJ, 198, 383\\
Lubow, S.H., Shu, F.H. 1976, ApJ, 207, L53\\
Ladous, C.1991, A\&A, 252, 100\\
Marsh, T.R., Horne, K., Rosen, S. 1991, ApJ, 366, 535
Marsh, T.R., 1999, MNRAS, 304, 443\\
Nagel T., Dreizler S., Werner K., 2003, in de Martino D., Kalytis R., Silvotti R., Solheim
   J., eds., White Dwarfs, Proc. XIII Workshop on White Dwarfs, Kluwer, pp. 357�
   358 \\
Nasser, M. R., Solheim, J.-E., \& Semionoff, D. A. 2001, A\&A, 373, 222
Nather, R.E., Robinson, E.L., Stover,R.J.,1981,ApJ,244,269 \\
Nather, E. 1985, in Interacting Binaries; Proceedings of the Advanced Study Institute,Cambridge, England, July 31-August 13, 1983,(Dordrecht:Reidel), 349 \\
Nelemans G., Portegies Zwart S.F., Verbunt F., Yungelson L.R., 2001a, A\&A, 368, 939 \\
Nelemans G., Steeghs D., Groot P.J., 2001b, MNRAS, 326, 621 \\
Nelemans G., Yungelson L.R., Portegies Zwart S.F., Verbunt F., 2001c, A\&A, 365, 491 \\
Nelemans, G.2005, ASPC, 330, 27\\
Paczynski B., 1967, Acta Astron., 17, 287 \\
Panei, J., Althaus, L.\& Benvenuto, O.2000, A\&Ap. 353, 970 \\
Ritter, H., \& Kolb, U.2003, A\&A, 404, 301\\
Roelofs G., Groot P., Steeghs D., Nelemans G., 2004, in Tovmassian G., Sion E., eds.,
Compact binaries in The Galaxy and beyond, volume 20 of RevMexAA (SC), p.
   254 \\
Roelofs et al., 2007a, ApJ, 666, 1174      \\
Roelofs, G.H.A., Groot, P.J., Nelemans, G., Marsh, T.R., Steeghs, D. 2007b, MNRAS, 379,
176 
Sion et al. 2006, ApJ, 636, L125\\
Sion et al. 2008, ApJ,104, 79 \\
Solheim, J.-E.2010, PASP, 122, 1133 \\
Solheim, J.-E., \& Yungelson, L.2005, 14th European Workshop on White Dwarfs, ASP Conference    Series, 334, 387 \\
Strohmayer T.E., 2004, ApJ, 608, L53  \\
Thorstensen, J.2005, AJ, 130, 759  \\
Urban, J., \& Sion, E.2006, ApJ, 642, 1029 \\
Warner, B.1995, Cataclysmic Variables, (Cambridge:CUP).  \\
Wood, M., et al. 1987, ApJ, 313, 757 \\


\begin{figure}
\plotone{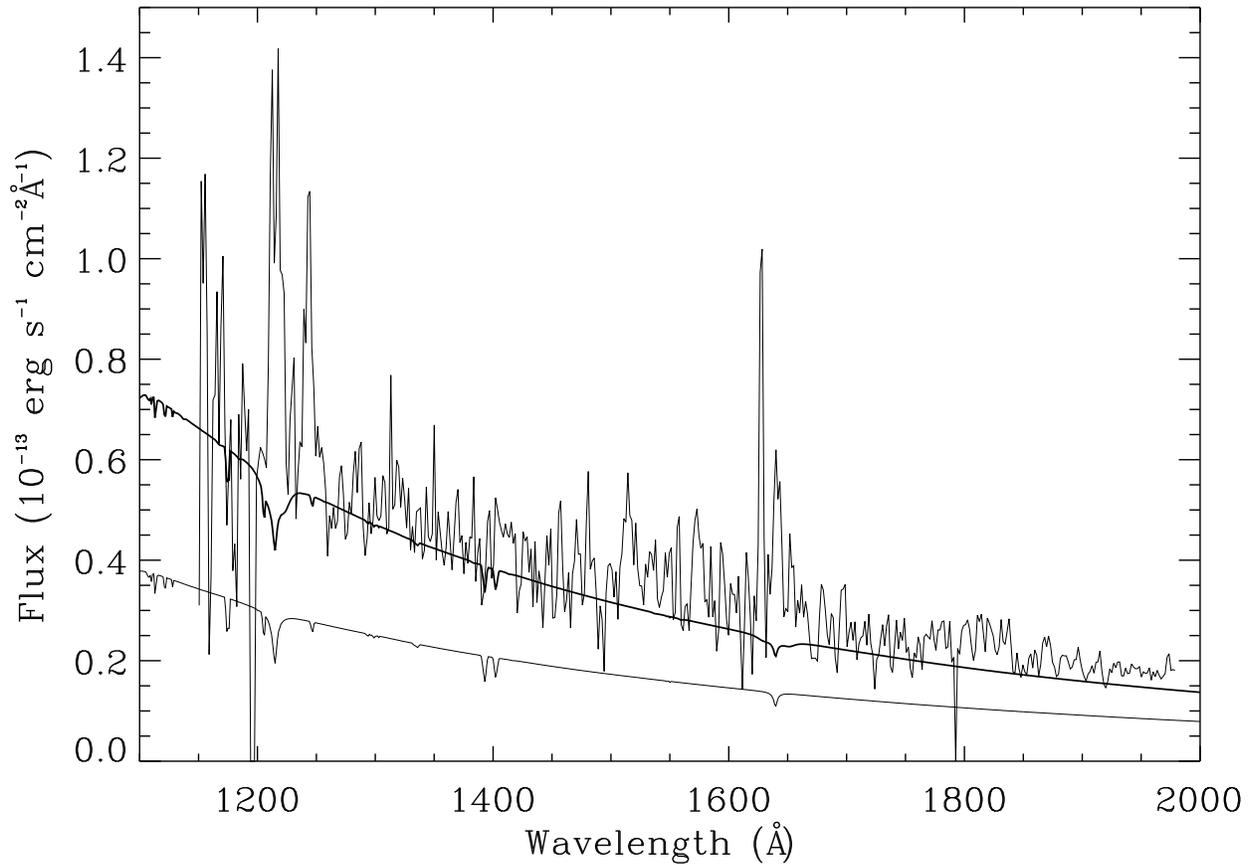}
\caption{The best-fitting accretion disk + white dwarf model to the IUE spectrum SWP30020
of the AM CVn system ES Ceti. The disk radius is $0.030R_{\odot}$ and the
 accretion rate $5 \times 10^{-10} M_{\odot}$/yr; see text for details.}
\end{figure}

\clearpage
\begin{figure}
\plotone{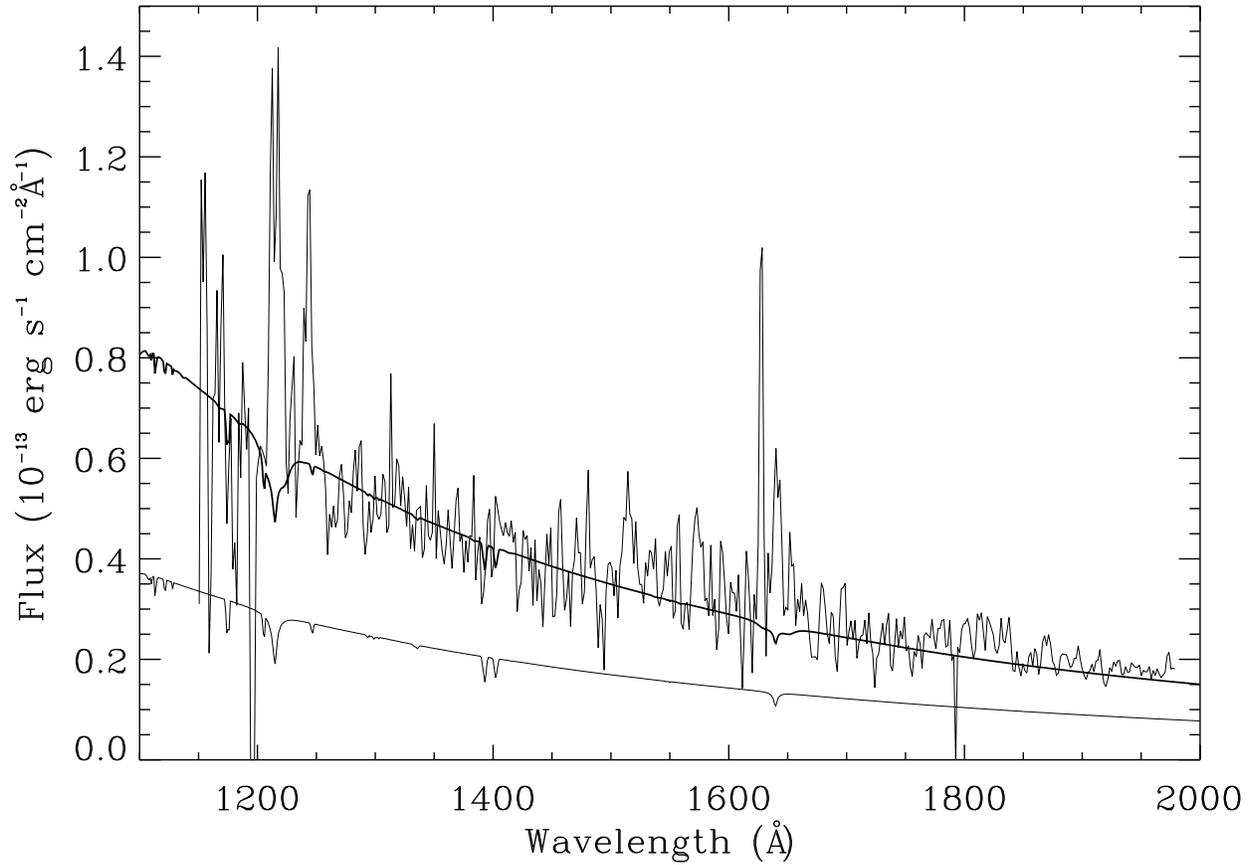}
\caption{The best-fitting accretion disk model to the IUE spectrum SWP30020 of the 
AM CVn system ES Ceti. The accretion disk has a disk radius $R_{disk} = 0.032 R_{\odot}$
and corresponds to \.{M}$ = 2.5\times 10^{-9}$ M$_{\odot}$ yr$^{-1}$; see the text for details.}
\end{figure}

\clearpage
\begin{figure}
\plotone{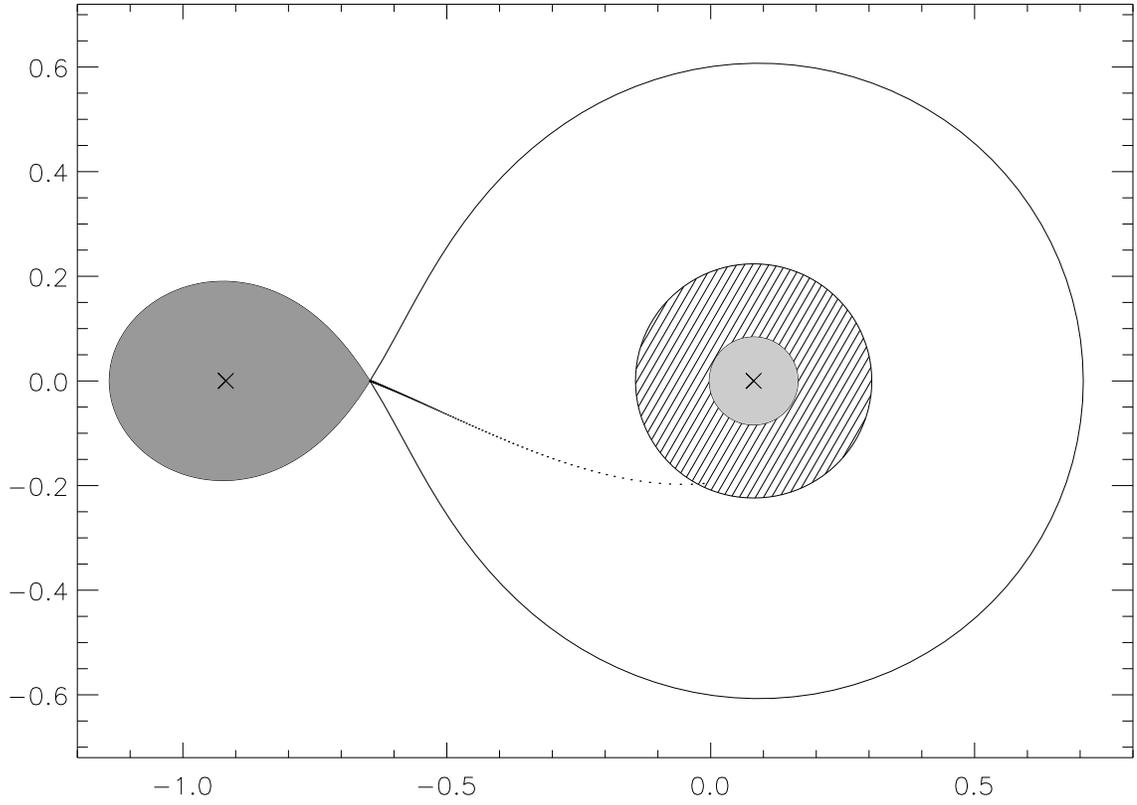}
\caption{An orbital plane view of the AM CVn system ES Ceti. The mass transfer stream intersects
the rim of the accretion disk. Continuation of the mass transfer stream trajectory misses the WD by a small amount.}
\end{figure}

\clearpage
\begin{figure}
\plotone{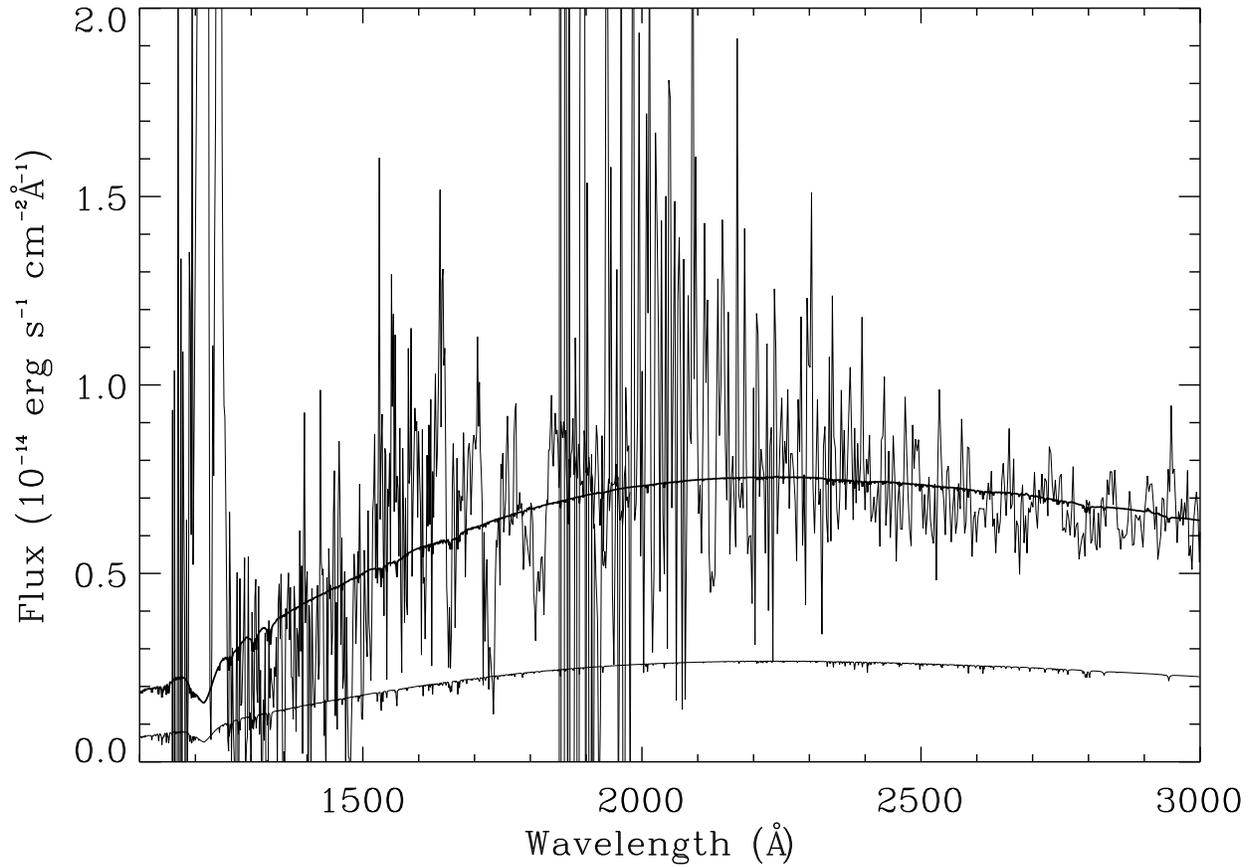}
\caption{The best-fitting accretion disk model to the IUE spectrum SWP46900 + LWP20586 of 
the AM CVn system GP Com. The accretion disk corresponds to \.{M}$ = 3.0-4.0\times 10^{-11}$ M$_{\odot}$ yr$^{-1}$, see text for details.}
\end{figure}

\clearpage
\begin{figure}
\plotone{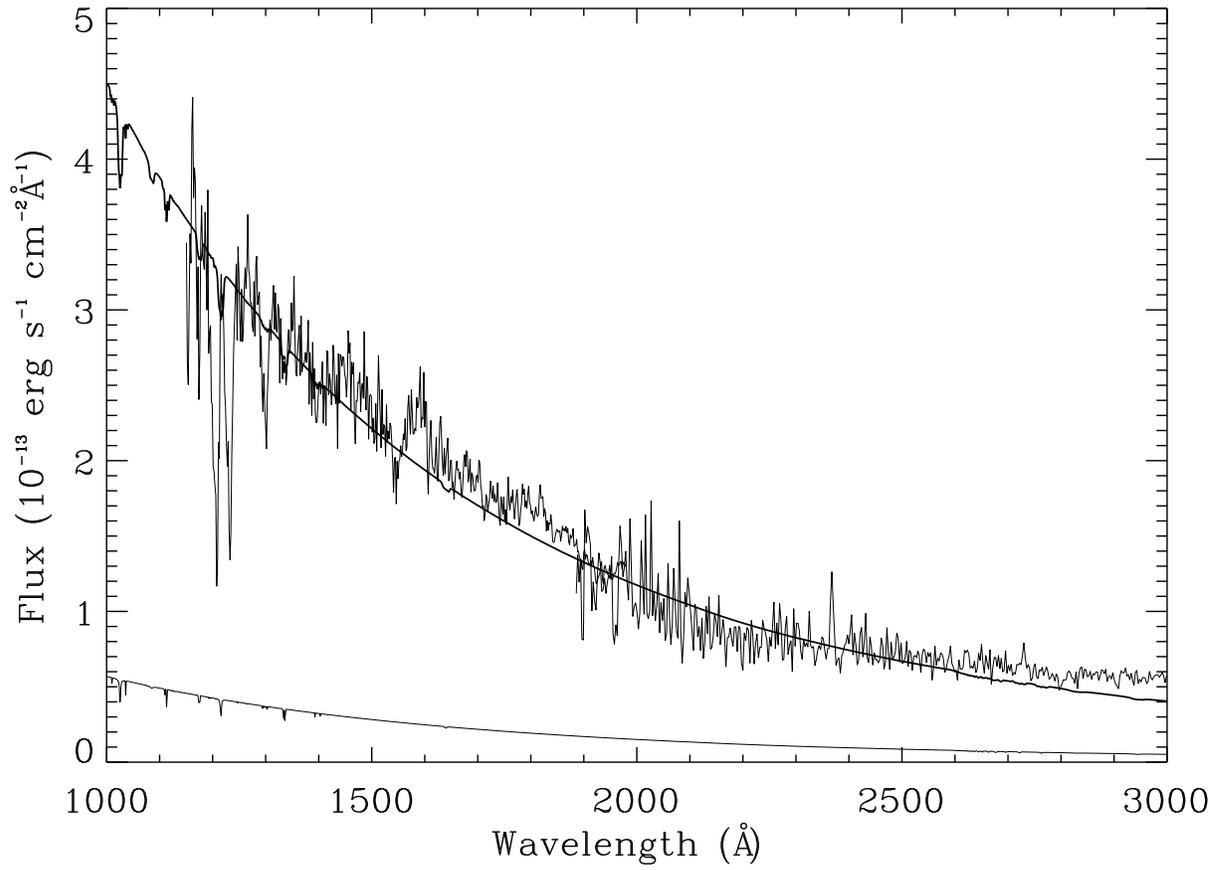}
\caption{The best-fitting accretion disk model to the IUE SWP spectrum only of the nova-like system HP Lib; see text for details.}
\end{figure}

\clearpage
\begin{figure}
\plotone{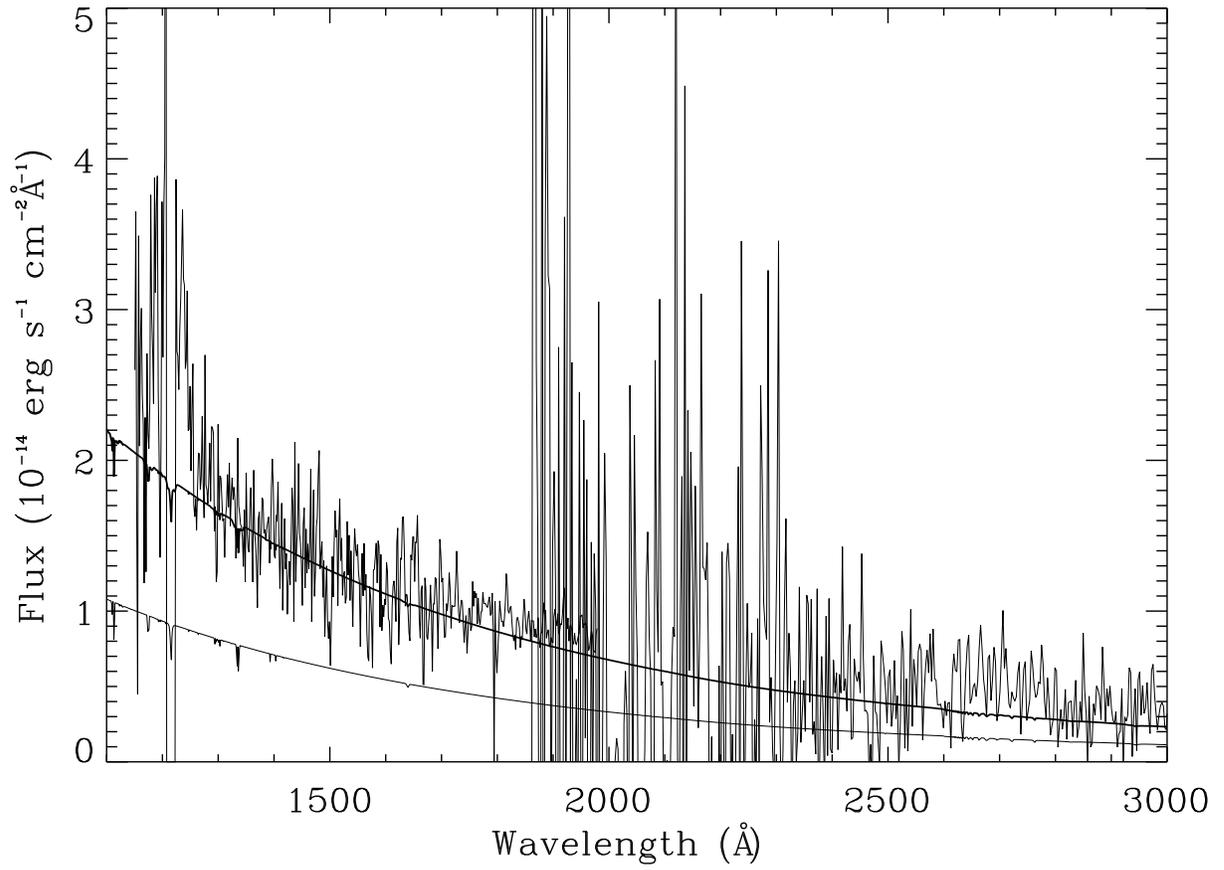}
\caption{The best-fitting accretion disk model to the IUE spectrum SWP33087 of CR Boo
during its low state. The accretion disk corresponds to \.{M}$ = 5.0\times 10^{-10}$M$_{\odot}$ yr$^{-1}$}

\end{figure}

\clearpage
\begin{figure}
\plotone{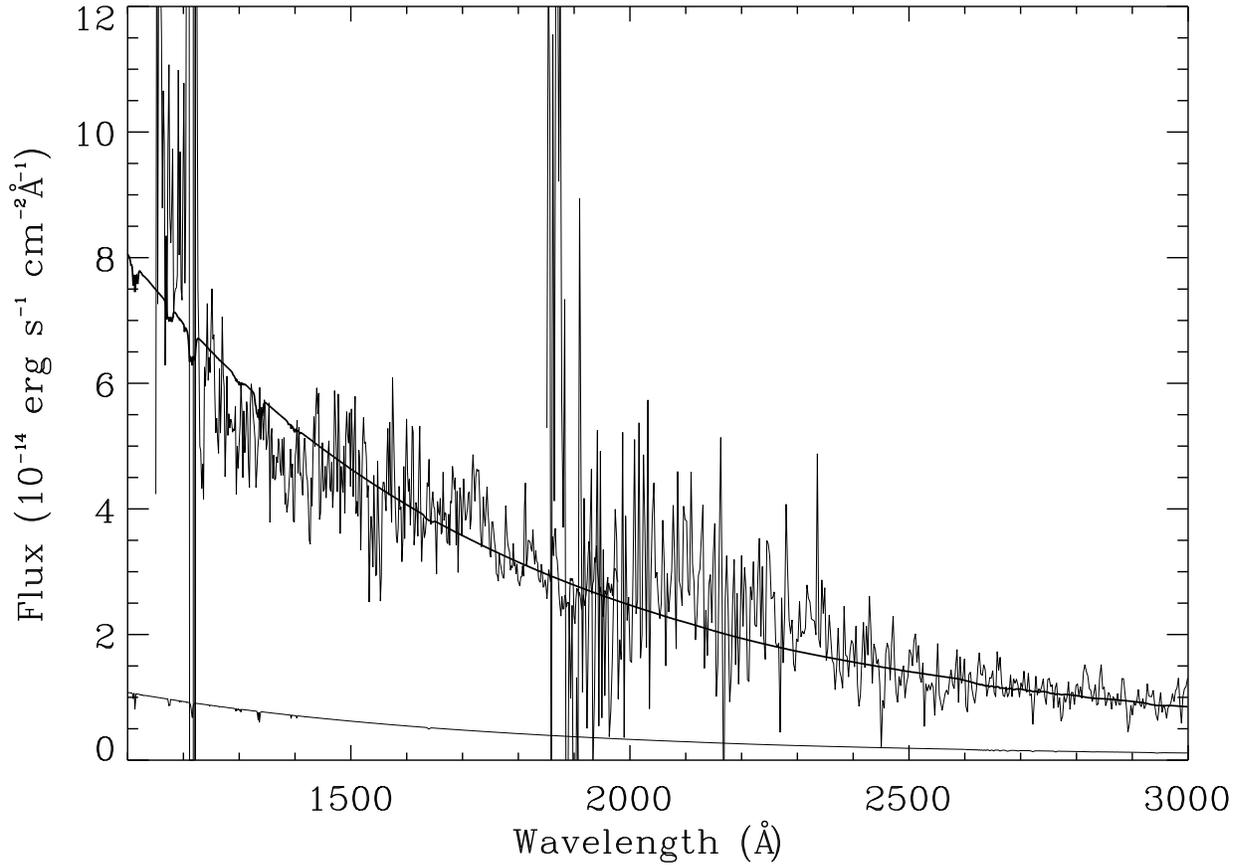}
\caption{The best-fitting accretion disk model to the IUE spectrum SWP33077 of CR Boo
during its high state; see the text for details.}
\end{figure}

\clearpage
\begin{figure}
\plotone{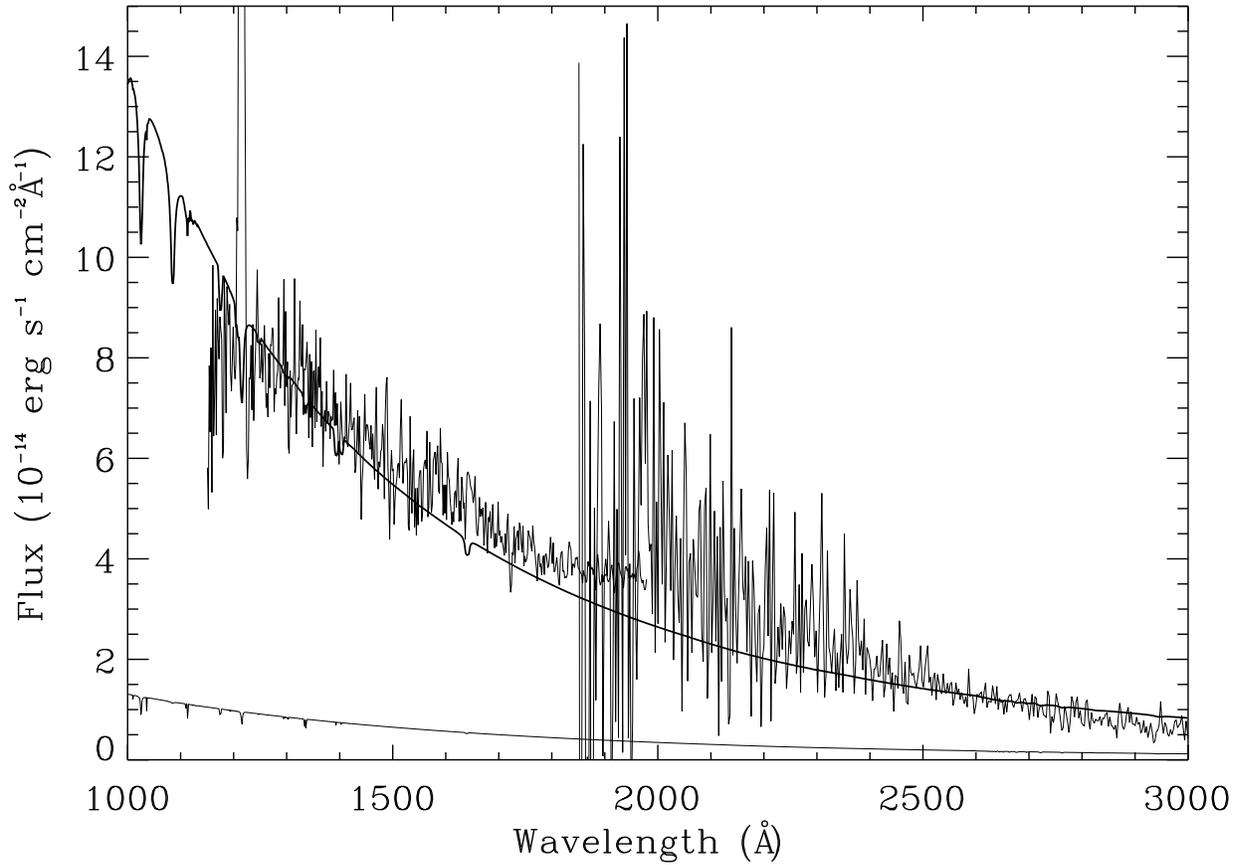}
\caption{The best-fitting accretion disk model to the IUE spectrum SWP44085 + LWP17482 of the nova-like system V803 Cen
during its low state. The accretion disk corresponds to \.{M}$ = 5.0\times 10^{-10}$ M$_{\odot}$ yr$^{-1}$, see text for details.}
\end{figure}

\clearpage
\begin{figure}
\plotone{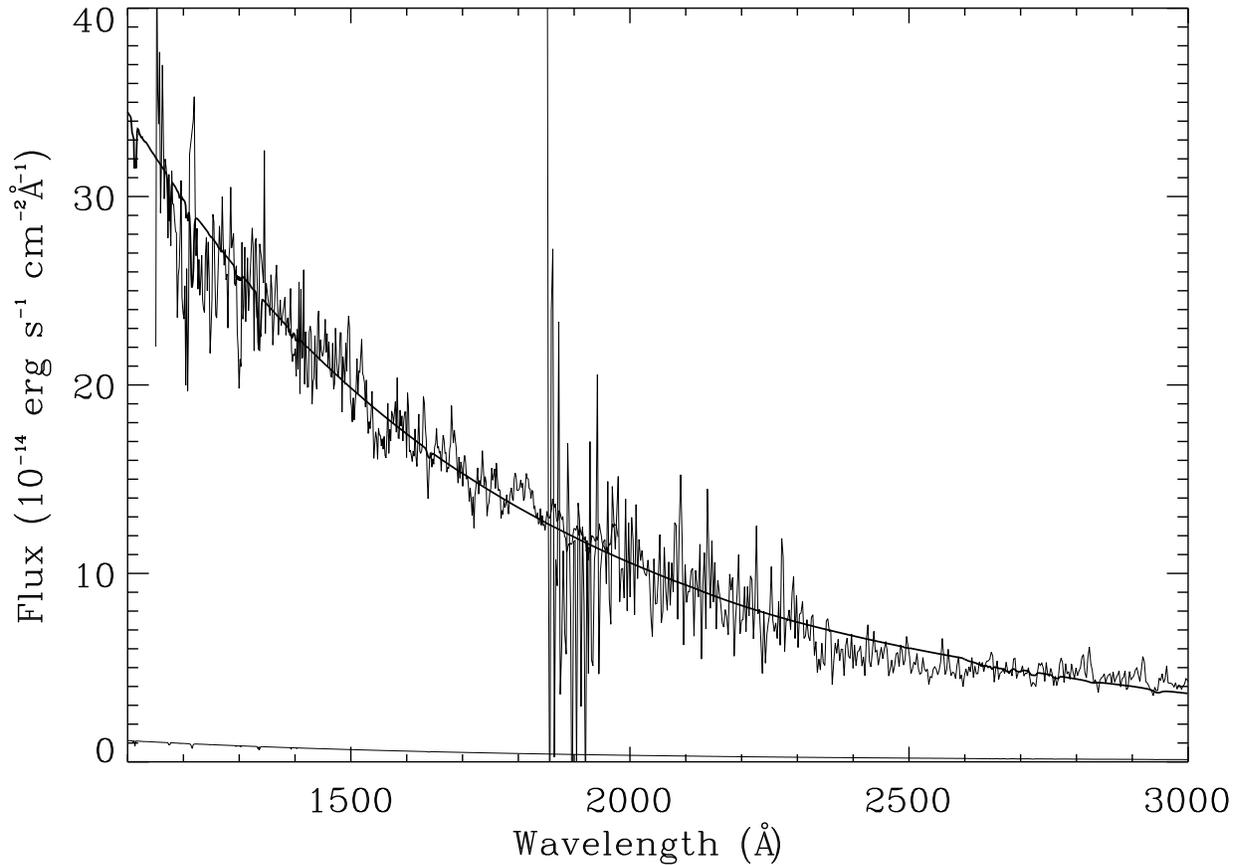}
\caption{The best-fitting accretion disk model to the IUE spectrum SWP38270 of the nova-like system V803 Cen
during its high state. The accretion disk corresponds to \.{M}$ = 3.0\times 10^{-10}$ M$_{\odot}$ yr$^{-1}$, see text for details.}
\end{figure}

\end{document}